\documentclass[fleqn,usenatbib,useAMS]{mnras}
 \usepackage{graphicx}
 \usepackage{amsmath}
 \usepackage{amssymb}
 \usepackage[T1]{fontenc}
 \usepackage{ae,aecompl}
 \usepackage{txfonts}
 \usepackage{enumerate}

 \title[Bimodal nodes: avoiding an unseen planet?]
       {Evidence for a possible bimodal distribution of the nodal distances
        of the extreme trans-Neptunian objects: avoiding a trans-Plutonian
        planet or just plain bias?}

 \author[C. de la Fuente Marcos and R. de la Fuente Marcos]
        {C.~de~la~Fuente~Marcos\thanks{E-mail: nbplanet@ucm.es}
         and
         R. de la Fuente Marcos \\
         Universidad Complutense de Madrid,
         Ciudad Universitaria, E-28040 Madrid, Spain}
 \date{Accepted 2017 June 22.
       Received 2017 June 21;
       in original form 2017 May 3}
 \pubyear{2017}
 
\begin{document}
  \label{firstpage}
  \pagerange{\pageref{firstpage}--\pageref{lastpage}}
  \maketitle

  \begin{abstract}
     It is a well-known fact that the presence of a massive perturber 
     interacting with a population of minor bodies following very eccentric 
     orbits can strongly affect the distribution of their nodal distances. 
     The details of this process have been explored numerically and its 
     outcome confirmed observationally in the case of Jupiter, where a 
     bimodal distribution of nodal distances of comets has been found. Here, 
     we show evidence for a possible bimodal distribution of the nodal 
     distances of the extreme trans-Neptunian objects (ETNOs) in the form of 
     a previously unnoticed correlation between nodal distance and orbital 
     inclination. This proposed correlation is unlikely to be the result of 
     observational bias as data for both large semimajor axis Centaurs and 
     comets fit well into the pattern found for the ETNOs, and all these 
     populations are subjected to similar background perturbations when 
     moving well away from the influence of the giant planets. The 
     correlation found is better understood if these objects tend to avoid a 
     putative planet with semimajor axis in the range of 300--400~au.
  \end{abstract}


  \section{Introduction}
     Beyond the orbit of Neptune, but moving in regions mostly unaffected by Galactic tides, lies a population of minor bodies following 
     very stretched elliptical paths. These large semimajor axis ($a>150$~au) and long perihelion distance ($q>30$~au) objects ---the 
     extreme trans-Neptunian objects or ETNOs (Trujillo \& Sheppard 2014)--- exhibit a number of unexpected orbital patterns that may be
     incompatible with the traditional eight-planets-only Solar system paradigm (see e.g. de la Fuente Marcos \& de la Fuente Marcos 2014; 
     Trujillo \& Sheppard 2014; Batygin \& Brown 2016a). 

     Peculiar orbital alignments can be caused by unseen massive perturbers in the framework of the trans-Plutonian planets hypothesis that 
     predicts that one (Trujillo \& Sheppard 2014; Batygin \& Brown 2016a; Brown \& Batygin 2016; Malhotra, Volk \& Wang 2016; Millholland 
     \& Laughlin 2017) or more (de la Fuente Marcos \& de la Fuente Marcos 2014, 2016a,b,c) yet-to-be-detected planetary bodies orbit the 
     Sun well beyond Pluto. This interpretation has been received with scepticism by some members of the scientific community who argue that 
     the currently available data are plagued with strong detection biases (Lawler et al. 2017; Shankman et al. 2017; Bannister et al. 
     2017); if observational biases are present, the data are inadequate to prove or disprove the presence of planets beyond Pluto. In stark 
     contrast, a mostly negligible role of detection biases is advocated by Trujillo \& Sheppard (2014), and by Batygin \& Brown (2016a) and 
     Brown \& Batygin (2016). Although such biases obviously exist as discussed by e.g. de la Fuente Marcos \& de la Fuente Marcos (2014), 
     they do not seem to work quite in the way one would had hoped; for example, most discoveries should have low orbital inclinations, but 
     this is not what is observed. 

     In the Solar system, Jupiter is the largest known perturber and a great deal can be learned about its impact on the orbital evolution 
     of minor bodies following very elongated paths by investigating particular examples. In this context, de la Fuente Marcos, de la Fuente 
     Marcos \& Aarseth (2015) have studied the case of its gravitational relationship with comet 96P/Machholz 1 as a relevant dynamical 
     analogue to better understand the behaviour of the ETNOs within the framework of the trans-Plutonian planets hypothesis. Their $N$-body 
     simulations show that Jupiter can send minor bodies into high-inclination or even retrograde orbits and also eject them from the Solar 
     system. Similar outcomes have been found within the context of the Planet Nine hypothesis (Batygin \& Brown 2016b; de la Fuente Marcos, 
     de la Fuente Marcos \& Aarseth 2016).

     However, Jupiter can affect the orbital architecture of the populations of minor bodies moving in highly eccentric paths in far more 
     subtle ways. Rickman, Valsecchi \& Froeschl{\'e} (2001) used numerical simulations to show that the values of the nodal distances of 
     comets with $a<1\,000$~au should follow a bimodal distribution. Here, we investigate the existence of a possible bimodal distribution 
     of the nodal distances of the ETNOs. This Letter is organized as follows. Section~2 reviews the case of Jupiter and the comets with 
     $a<1\,000$~au. The case of the ETNOs is explored in Section~3. The analysis is further extended to large semimajor axis or extreme 
     Centaurs and comets in Section~4. In Section~5, we study the statistical significance of our findings. Results are discussed in 
     Section~6, and conclusions are summarized in Section~7.

  \section{Jupiter and the comets with $\mathbf{\lowercase{a}<1\,000}$~\lowercase{au}}
     Rickman et al. (2001) investigated numerically the delivery of comets from the Oort cloud to the inner Solar system. In their fig.~6, 
     they show that for comets captured in an orbit with $a<1\,000$~au, the distribution of nodal distances (for the node closest to 
     Jupiter's orbit), $r_{\rm n}$, should be bimodal with a primary peak near the present-day value of the semimajor axis of Jupiter 
     (5.2~au) and a secondary peak in the neighbourhood of the Earth. For a direct or prograde elliptical orbit, the distance from the Sun 
     to the nodes can be written as 
     \begin{equation}
        r_{\rm n} = \frac{a\ (1 - e^2)}{1\pm{e}\cos\omega}\,, \label{nodeseq}
     \end{equation}
     where $e$ is the eccentricity, $\omega$ is the argument of perihelion and the `+' sign is for the ascending node (where the orbit 
     passes from south to north, crossing the ecliptic) and the `$-$' sign is for the descending node ---for a retrograde orbit (inclination, 
     $i>90\degr$) the signs are switched. Rickman et al. (2001) showed that the primary peak is associated with captures due to close 
     encounters with Jupiter, while the secondary peak represents captures due to indirect perturbations because comets with short 
     perihelion distances experience the largest perturbation as they speed up closer to the Sun. In addition, the strength of mean-motion 
     resonances (in this case with Jupiter) is proportional to the eccentricity of the perturbed body (e.g. Nesvorn{\'y} \& Roig 2001).

\begin{figure}
        \centering
         \includegraphics[width=\linewidth]{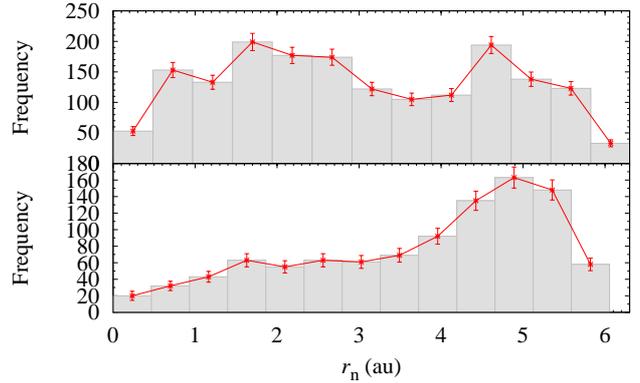}
         \caption{The distribution of the nodal distances of observed comets with $a<1\,000$~au. Top panel, ascending and descending nodes 
                  (IQR=2.91~au, $n$=1\,716); bottom panel, node closest to Jupiter (IQR=2.33~au, $n$=1\,002).  
                 }
         \label{jupc}
      \end{figure}

     As of 2017 June 5, the number of comets with $a<1\,000$~au stands at 1\,120 ---here and in other sections, we use data provided by Jet 
     Propulsion Laboratory's Solar System Dynamics Group Small-Body Database (JPL's SSDG SBDB, Giorgini 
     2011).\footnote{\url{https://ssd.jpl.nasa.gov/sbdb.cgi}} Fig.~\ref{jupc}, bottom panel, is equivalent to fig.~6 in Rickman et al. 
     (2001) but for real comets. In the histograms, we adopt Poisson statistics ($\sigma=\sqrt{n}$) to compute the error bars ---applying 
     the approximation given by Gehrels (1986) when $n<21$, $\sigma \sim 1 + \sqrt{0.75 + n}$--- and the bin width has been found using the 
     Freedman-Diaconis rule (Freedman \& Diaconis 1981), i.e. $2\ {\rm IQR}\ n^{-1/3}$, where IQR is the interquartile range and $n$ is the 
     number of data points. 

     Using observational data, we confirm that the distribution of nodal distances is indeed bimodal. Fig.~\ref{jupc}, top panel, is very 
     similar to fig.~8, left-hand panel, in Rickman \& Froeschl{\'e} (1988) that shows the distribution of nodal distances for short-period 
     comets of the Jupiter family and includes both ascending and descending nodes. We observe that nodes tend to avoid the neighbourhood of 
     the 3:2 and/or the 2:1 mean-motion resonances with Jupiter, but it is unclear why is this happening. These results are statistically 
     significant ($>8\sigma$), if we consider the associated Poisson uncertainties, and they are unlikely to be affected by observational 
     biases or completeness of the sample issues. Given this finding, it is not unreasonable to assume that massive perturbers, if present 
     in the outer Solar system, should induce a similar behaviour on the populations of eccentric small bodies existing there.
%
%
%
%
%
%
      \begin{table*}
        \centering
        \fontsize{8}{11pt}\selectfont
        \tabcolsep 0.05truecm
        \caption{Orbital elements (heliocentric and barycentric) and nodal distances of the ETNOs with 1$\sigma$ uncertainties. Data include 
                 the heliocentric and barycentric semimajor axis, $a$ and $a_{\rm b}$, eccentricity, $e$ and $e_{\rm b}$, inclination, $i$
                 and $i_{\rm b}$, longitude of the ascending node, $\Omega$ and $\Omega_{\rm b}$, argument of perihelion, $\omega$ and 
                 $\omega_{\rm b}$, and their respective standard deviations, $\sigma_a$, $\sigma_e$, $\sigma_i$, $\sigma_{\Omega}$ and 
                 $\sigma_{\omega}$; the computed heliocentric nodal distances, $r_{+}$ and $r_{-}$, and their barycentric counterparts,
                 $r_{{\rm b}+}$ and $r_{{\rm b}-}$, with their respective standard deviations (the `+' sign is for the ascending node and 
                 the `$-$' sign is for the descending node). The orbital solutions have been computed at epoch JD 2458000.5 that corresponds 
                 to 00:00:00.000 TDB on 2017 September 4, J2000.0 ecliptic and equinox. Source: JPL's SSDG SBDB.
                }
        \resizebox{\linewidth}{0.22\linewidth}{
        \begin{tabular}{lrrrrrrrrrrrrrrrcccc}
          \hline
             Object          & $a$ (au) & $\sigma_a$ (au) & $a_{\rm b}$ (au) & $e$ & $\sigma_e$ & $e_{\rm b}$ & $i$ (\degr) &$\sigma_i$ (\degr) & $i_{\rm b}$ (\degr)
                             & $\Omega$ (\degr) & $\sigma_{\Omega}$ (\degr) & $\Omega_{\rm b}$ (\degr) & $\omega$ (\degr) & $\sigma_{\omega}$ (\degr) 
                             & $\omega_{\rm b}$ (\degr) & $r_{+}$ (au)      & $r_{-}$ (au)     & $r_{{\rm b}+}$ (au) & $r_{{\rm b}-}$ (au) \\ 
          \hline
             82158           & 227.23    &  0.07      & 215.49           & 0.84914 & 0.00004    & 0.84106     & 30.762880   & 0.000053    & 30.800391
                             & 179.31043        & 0.00006                   & 179.35869                &   7.1638         &  0.0011 	
                             &   6.8733                 & 34.404$\pm$0.014  & 402.49$\pm$0.12  & 34.364$\pm$0.014    & 382.19$\pm$0.12     \\ 
             Sedna           & 487.77    &  0.62      & 507.42           & 0.84409 & 0.00021    & 0.84987     & 11.929003   & 0.000009    & 11.928559
                             & 144.45959        & 0.00147                   & 144.40308                & 311.6130         &  0.0115    
                             & 311.3066                 & 89.864$\pm$0.169  & 319.13$\pm$0.52  & 90.275$\pm$0.166    & 320.99$\pm$0.51     \\
            148209           & 222.92    &  0.65      & 221.98           & 0.80139 & 0.00057    & 0.80123     & 22.756428   & 0.000585    & 22.755916
                             & 128.29160        & 0.00030                   & 128.28584                & 317.0376         &  0.0117    
                             & 316.6901                 & 50.273$\pm$0.208  & 192.86$\pm$0.65  & 50.205$\pm$0.208    & 190.60$\pm$0.64     \\
            445473           & 150.48    &  0.02      & 153.43           & 0.77185 & 0.00004    & 0.77611     &  4.510987   & 0.000024    &  4.510524
                             & 117.39460        & 0.00147                   & 117.39674                & 313.8822         &  0.0028 
                             & 313.7247                 & 39.629$\pm$0.009  & 130.83$\pm$0.03  & 39.710$\pm$0.009    & 131.62$\pm$0.02     \\
            474640           & 315.40    &  1.75      & 326.91           & 0.84999 & 0.00081    & 0.85525     & 25.592704   & 0.000317    & 25.547995
                             &  65.98210        & 0.00053                   &  66.02214                & 326.9906         &  0.0093  
                             & 326.9871                 & 51.104$\pm$0.396  & 304.75$\pm$1.85  & 51.126$\pm$0.394    & 310.40$\pm$1.86     \\
       	     2002 GB$_{32}$  & 219.01    &  0.78      & 206.71           & 0.83860 & 0.00056    & 0.82903     & 14.175870   & 0.000310    & 14.192093
                             & 176.98897        & 0.00044                   & 177.04362                &  37.1585         &  0.0047 
                             &  37.0472                 & 38.955$\pm$0.193  & 195.95$\pm$0.79  & 38.900$\pm$0.192    & 191.06$\pm$0.78     \\
             2003 HB$_{57}$  & 166.14    &  0.71      & 159.59           & 0.77061 & 0.00094    & 0.76128     & 15.472470   & 0.001354    & 15.500156
                             & 197.82250        & 0.00043                   & 197.87105                &  11.0091         &  0.0631 
                             &  10.8330                 & 38.417$\pm$0.232  & 277.04$\pm$1.20  & 38.394$\pm$0.229    & 265.97$\pm$1.19     \\
             2003 SS$_{422}$ & 199.47    & 148.31     & 203.26           & 0.80296 & 0.16161    & 0.80657     & 16.811966   & 0.147140    & 16.781796
                             & 151.08067        & 0.17403                   & 151.04186                & 211.7279         & 43.1730 	
                             & 211.5975                 & 224$\pm$1733      & 42$\pm$69        & 227$\pm$1853        & 42$\pm$77          \\
             2005 RH$_{52}$  & 152.00    &  0.26      &	153.65           & 0.74365 & 0.00040    & 0.74620     & 20.445528   & 0.000718    & 20.446049
                             & 306.09328        & 0.00172                   & 306.11117                &  32.3110         &  0.0619   
                             &  32.5448                 & 41.720$\pm$0.096  & 182.88$\pm$0.38  & 41.802$\pm$0.094    & 183.56$\pm$0.38     \\
             2007 TG$_{422}$ & 471.70    &  0.42      & 502.81           & 0.92462 & 0.00007    & 0.92927     & 18.603413   & 0.000075    & 18.595308
                             & 112.89216        & 0.00034                   & 112.91072                & 285.6562         &  0.0034 
                             & 285.6605                 & 54.770$\pm$0.068  &  91.19$\pm$0.11  & 54.850$\pm$0.070    &  91.58$\pm$0.11     \\
             2007 VJ$_{305}$ & 188.02    &  0.16      & 192.17           & 0.81298 & 0.00016    & 0.81690     & 12.004643   & 0.000171    & 11.983580
                             &  24.38369        & 0.00008                   &  24.38259                & 338.1883         &  0.0038 
                             & 338.3541                 & 36.331$\pm$0.045  & 259.98$\pm$0.24  & 36.337$\pm$0.044    & 265.59$\pm$0.23     \\
             2010 GB$_{174}$ & 363.66    & 25.46      & 351.38           & 0.86583 & 0.01021    & 0.86181     & 21.560511   & 0.005128    & 21.562666
                             & 130.71336        & 0.01973                   & 130.71530                & 347.7672         &  0.3671 
                             & 347.2366                 & 49.311$\pm$5.080  & 591.81$\pm$42.26 & 49.119$\pm$5.091    & 566.85$\pm$42.17    \\
             2012 VP$_{113}$ & 255.76    &  1.34      & 262.07           & 0.68525 & 0.00195    & 0.69274     & 24.085640   & 0.002320    & 24.052058
                             &  90.73148        & 0.00562                   &  90.80272                & 293.8367         &  0.3765 
                             & 293.9250                 & 106.24$\pm$0.89   & 187.62$\pm$1.64  & 106.41$\pm$0.88     & 189.56$\pm$1.66     \\
             2013 FS$_{28}$  & 196.70    & 98.78      & 191.76           & 0.82439 & 0.09765    & 0.82134     & 13.006215   & 0.024737    & 13.068231
                             & 204.67337        & 0.01617                   & 204.63813                & 101.5395         &  2.4474
                             & 102.1765                 &  75.47$\pm$55.58  &  54.10$\pm$41.92 &  75.47$\pm$53.68    &  53.18$\pm$39.62    \\
             2013 FT$_{28}$  & 312.28    & 10.53      & 294.52           & 0.86051 & 0.00505    & 0.85239     & 17.329026   & 0.003416    & 17.375249
                             & 217.78017        & 0.00483                   & 217.72271                &  40.2649         &  0.1672
                             &  40.6969                 &  48.92$\pm$2.47   & 236.02$\pm$9.87  &  48.92$\pm$2.46     & 227.65$\pm$9.62     \\
             2013 GP$_{136}$ & 154.27    &  0.82      & 149.71           & 0.73359 & 0.00168    & 0.72571     & 33.466607   & 0.001909    & 33.538942
                             & 210.70939        & 0.00023                   & 210.72727                &  42.2113         &  0.1643 
                             &  42.4635                 &  46.16$\pm$0.37   & 156.02$\pm$1.04  &  46.16$\pm$0.37     & 152.52$\pm$1.04     \\
             2013 RF$_{98}$  & 349.23    & 11.73      & 363.87           & 0.89667 & 0.00358    & 0.90080     & 29.579219   & 0.003374    & 29.538373
                             &  67.58666        & 0.00532                   &  67.63560                & 311.7287         &  0.6725 
                             & 311.7566                 &  42.86$\pm$2.04   & 169.77$\pm$8.07  &  42.89$\pm$2.07     & 171.49$\pm$8.07     \\
             2013 SY$_{99}$  & 672.89    & 21.43      & 729.24           & 0.92578 & 0.00245    & 0.93147     &  4.233857   & 0.001201    &  4.225428
                             &  29.47329        & 0.00519                   &  29.50927                &  32.3248         &  0.1138
                             &  32.1410                 &  53.96$\pm$2.46   & 441.81$\pm$17.31 &  53.96$\pm$2.50     & 456.82$\pm$17.79    \\
             2013 UH$_{15}$  & 170.66    &  8.30      & 173.75           & 0.79524 & 0.01131    & 0.79846     & 26.127110   & 0.005795    & 26.080631
                             & 176.60152        & 0.00721                   & 176.54233                & 283.0936         &  0.2724
                             & 282.8653                 &  53.16$\pm$3.74   &  76.52$\pm$5.12  &  53.47$\pm$3.75     &  76.60$\pm$5.09     \\
             2014 FE$_{72}$  & 1836.42   & 2066.40    & 1559.28          & 0.98020 & 0.02248    & 0.97680     & 20.616558   & 0.008942    & 20.637561
                             & 336.80375        & 0.01621                   & 336.83831                & 134.3877         &  0.2131
                             & 133.9213                 &  229$\pm$475      &     43$\pm$94    &    222$\pm$421      &     43$\pm$85      \\
             2014 SR$_{349}$ & 294.06    & 18.29      & 298.50           & 0.83813 & 0.01107    & 0.84073     & 17.984844   & 0.002072    & 17.968246
                             &  34.75185        & 0.01736                   &  34.88438                & 341.2503         &  0.6557
                             & 341.2593                 &  48.78$\pm$4.64   & 424.01$\pm$27.78 &  48.72$\pm$4.65     & 429.31$\pm$27.93    \\
             2015 SO$_{20}$  & 161.62    & 0.04       & 164.79           & 0.79481 & 0.00005    & 0.79871     & 23.451236   & 0.000136    & 23.410786
                             &  33.61877        & 0.00009                   &  33.63407                & 354.8049         &  0.0063 
                             & 354.8329                 & 33.225$\pm$0.012  & 285.54$\pm$0.07  & 33.229$\pm$0.012    & 291.70$\pm$0.07     \\
          \hline
        \end{tabular}
        }
        \label{etnos}
      \end{table*}
%
%
  \section{A bimodal distribution for the ETNOs?}
\begin{figure}
        \centering
         \includegraphics[width=\linewidth]{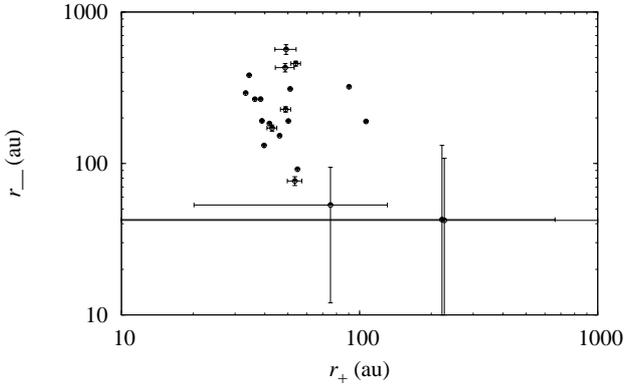}
         \caption{Distance to the descending node as a function of the distance to the ascending node for the 22 known ETNOs. The nodal 
                  distances have been found using the barycentric elements and the error bars have been computed using the uncertainties
                  in Table~\ref{etnos} (see the text for details).   
                 }
         \label{nodes}
      \end{figure}

     As of 2017 June 5, the number of trans-Neptunian objects with $a>150$~au and $q>30$~au stands at 22 (see Table~\ref{etnos}). It is 
     therefore not advisable to try to produce a histogram similar to those in Fig.~\ref{jupc} only for this population. The uncertainties 
     for 2003~SS$_{422}$ and 2014~FE$_{72}$ are the largest by far, and the quality of their orbital solutions does not allow us to reach 
     any robust conclusion on the location of their nodes. The remaining data appear to be reasonably good and show that ascending nodes 
     tend to be systematically located at barycentric distances $<200$~au, which was expected as most objects in Table~\ref{etnos} have 
     values of the argument of perihelion close to 0\degr. Surprisingly, the only examples of ETNOs with ascending nodes beyond 200~au 
     correspond to the most uncertain orbits so their nominal locations may be incorrect (see Fig.~\ref{nodes} and Table~\ref{etnos}). 

     Fig.~\ref{nodes} shows the distance to the descending node as a function of the distance to the ascending node of the ETNOs computed 
     using barycentric elements (see Table~\ref{etnos}). The distances have been calculated (averages and standard deviations, see e.g. Wall 
     \& Jenkins 2012) using 2\,500 instances of the set of parameters $a$, $e$ and $\omega$ for each object and applying 
     Equation~(\ref{nodeseq}). For example, new values of $a$ for a given ETNO have been found using the expression 
     $\langle{a}\rangle + \sigma_{a}\,r_{\rm i}$, where $\langle{a}\rangle$ is the nominal value of the semimajor axis from 
     Table~\ref{etnos}, $\sigma_{a}$ is its associated standard deviation, also from Table~\ref{etnos}, and $r_{\rm i}$ is a (pseudo) random 
     number with standard normal distribution. The Box-Muller method (Box \& Muller 1958; Press et al. 2007) has been applied to generate 
     the random numbers from a standard normal distribution with mean 0 and standard deviation 1. As 64 bits computers were used to complete 
     the calculations, the smallest non-zero number is $2^{-64}$, and the Box-Muller method will not produce random variables more than 9.42 
     standard deviations from the mean. This fact should not have any adverse impact on our results, the standard deviations of the nodal 
     distances in Table~\ref{etnos} are expected to be correct. Although it can be argued that the present orbital solutions of many ETNOs 
     are too uncertain to use the data for this type of analysis, we believe that accounting for the role of the errors as we have done here 
     greatly minimizes this issue.

\begin{figure}
        \centering
         \includegraphics[width=\linewidth]{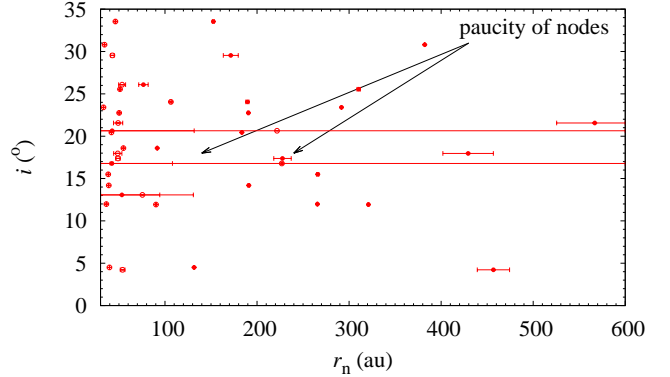}
         \caption{Orbital inclination as a function of the nodal distances. Nodal distances and error bars have been computed as in 
                  Fig.~\ref{nodes}. Here and in Fig.~\ref{inodesALL}, the arrows point out the regions where the paucities are located (see 
                  the text for details).
                 }
         \label{inodes}
      \end{figure}

     In addition, Equation~(\ref{nodeseq}) shows that $r_{\rm n}$ does not depend on the orbital inclination and, in absence of significant 
     direct perturbations or resonances, one may assume that the values of nodal distances and inclinations must be uncorrelated. 
     Fig.~\ref{inodes} shows the orbital inclination as a function of the nodal distances for the same sample. Nodal distances in the range 
     (200, 250)~au seem to be avoided and the few present have associated values of the inclination within a rather narrow range; a paucity 
     of nodes in the range (100, 150)~au does not show any obvious correlation with $i$. Although based on small-number statistics, the 
     trend resembles that of the comets with $a<1\,000$~au discussed in the previous section. If this correlation is real, it should also be 
     observed among other populations that experience similar background perturbations during most of their orbital periods when moving well 
     away from the influence of the giant planets. Such populations include the extreme Centaurs ($a>150$~au but $q<30$~au) and comets with 
     $a>150$~au and $e<1$.
%
%
%
%
%
%
%
%
%
%
      \begin{table*}
        \centering
        \fontsize{8}{11pt}\selectfont
        \tabcolsep 0.05truecm
        \caption{As Table~\ref{etnos} but for the known extreme Centaurs.
                }
        \begin{tabular}{lrrrrrrrrrrcc}
          \hline
             Object          & $a$ (au)  & $\sigma_a$  (au) & $e$      & $\sigma_e$ & $i$ (\degr) &  $\sigma_i$ (\degr) & $\Omega$ (\degr) & $\sigma_{\Omega}$ (\degr) 
                             & $\omega$ (\degr) & $\sigma_{\omega}$ (\degr) & $r_{+}$ (au)      & $r_{-}$ (au)     \\ 
          \hline
             87269           & 507.07    &  1.91      & 0.959083 & 0.000153   & 20.071069   & 0.000315    & 142.336335       & 0.000279  
                             & 212.24223        &   0.00360                 & 215.28$\pm$1.03   &  22.44$\pm$0.12  \\ 	
             308933          & 785.55    &  0.94      & 0.969269 & 0.000037   & 19.498353   & 0.000056    & 197.341739       & 0.000094  
                             & 122.27520        &   0.00068                 &  98.54$\pm$0.16   &  31.33$\pm$0.05  \\ 	
             336756          & 308.92    &  0.10      & 0.969570 & 0.000010   & 140.784619  & 0.000021    & 136.171903       & 0.000012  
                             & 132.76335        &   0.00017                 &  11.165$\pm$0.005 &  54.19$\pm$0.02  \\ 	
             418993          & 355.27    &  0.11      & 0.969048 & 0.000009   & 68.079230   & 0.000015    & 220.222727       & 0.000014  
                             & 128.61714        &   0.00008                 &  54.79$\pm$0.02   & 13.492$\pm$0.006 \\ 	
             468861          & 173.49    &  0.16      & 0.949692 & 0.000045   & 125.361075  & 0.000093    & 276.009114       & 0.000163  
                             & 153.18391        &   0.00201                 & 9.211$\pm$0.012   & 111.63$\pm$0.12  \\ 	
             469750          & 170.92    &  0.06      & 0.828927 & 0.000063   &  6.189265   & 0.000186    & 192.460952       & 0.000540  
                             & 227.99820        &   0.00242                 & 120.09$\pm$0.05   &  34.40$\pm$0.02  \\ 	
             1996 PW         & 251.90    &  0.32      & 0.990001 & 0.000013   & 29.796991   & 0.000065    & 144.539029       & 0.000030  
                             & 181.60094        &   0.00043                 & 482.64$\pm$0.61   &  2.519$\pm$0.005 \\ 	
             2002 RN$_{109}$ & 578.90    & 25.88      & 0.995329 & 0.000209   & 57.926426   & 0.004905    & 170.502698       & 0.000057  
                             & 212.38942        &   0.00580                 &  33.83$\pm$2.14   &   2.93$\pm$0.19  \\ 	
             2005 VX$_{3}$   & 1103.63   & 240.08     & 0.996267 & 0.000812   & 112.273081  & 0.007645    & 255.317842       & 0.001722  
                             & 196.49382        &   0.01034                 &   4.21$\pm$1.31   & 183.86$\pm$54.79 \\ 	
             2006 UL$_{321}$ & 260.76    &   $-$      & 0.909904 &   $-$      & 37.366730   &   $-$       & 342.920200       &   $-$      
                             & 353.79724        &     $-$                   & 23.56             & 470.23           \\ 	
             2007 DA$_{61}$  & 495.17    & 64.93      & 0.994602 & 0.000707   & 76.767820   & 0.009274    & 145.880900       & 0.002377  
                             & 349.75417        &   0.01251                 &   2.69$\pm$0.49   & 250.80$\pm$40.53 \\ 	
             2010 BK$_{118}$ & 432.62    &  0.30      & 0.985917 & 0.000010   & 143.913110  & 0.000023    & 176.002953       & 0.000021  
                             & 178.93966        &   0.00011                 &  6.093$\pm$0.006  & 848.97$\pm$0.59  \\ 	
             2010 GW$_{147}$ & 166.15    &  0.38      & 0.967601 & 0.000073   & 99.718297   & 0.000088    & 313.305837       & 0.000169  
                             &  50.05406        &   0.00167                 &  27.97$\pm$0.09   &   6.53$\pm$0.02  \\ 	
             2011 OR$_{17}$  & 256.88    &  0.13      & 0.987921 & 0.000006   & 110.442034  & 0.000026    & 271.490201       & 0.000036  
                             &  13.97512        &   0.00011                 & 149.28$\pm$0.10   &  3.149$\pm$0.002 \\ 	
             2012 DR$_{30}$  & 1595.55   &  3.72      & 0.990873 & 0.000021   & 77.966456   & 0.000046    & 341.415546       & 0.000023  
                             & 195.53157        &   0.00029                 & 639.86$\pm$1.88   &  14.83$\pm$0.05  \\ 	
             2012 GU$_{11}$  & 183.93    &  0.07      & 0.900823 & 0.000036   & 10.751141   & 0.000041    & 144.788872       & 0.000234  
                             &   6.55747        &   0.00101                 & 18.298$\pm$0.010   & 330.00$\pm$0.12  \\ 	
             2012 KA$_{51}$  & 224.45    & 2030.00    & 0.978126 & 0.197340   & 70.661491   & 0.251860    & 344.953286       & 0.484370  
                             &  95.03797        &   5.57050                 & 11$\pm$906    &      9$\pm$754   \\ 	
             2013 AZ$_{60}$  & 562.45    &  0.36      & 0.985918 & 0.000009   & 16.531904   & 0.000014    & 349.201540       & 0.000028  
                             & 158.41894        &   0.00006                 & 189.06$\pm$0.16   &   8.206$\pm$0.007 \\ 	
             2013 BL$_{76}$  & 1080.07   &  2.77      & 0.992258 & 0.000020   & 98.613012   & 0.000053    & 180.196765       & 0.000012  
                             & 165.95965        &   0.00042                 &   8.49$\pm$0.03   & 445.61$\pm$1.45  \\ 	
             2014 GR$_{53}$  & 221.56    &  0.12      & 0.897855 & 0.000053   & 41.992112   & 0.000190    &  52.559667       & 0.000087  
                             & 177.79558        &   0.00096                 & 417.77$\pm$0.22   &  22.64$\pm$0.02  \\ 	
             2014 LM$_{28}$  & 268.41    &  0.76      & 0.937525 & 0.000183   & 84.747713   & 0.000632    & 246.186279       & 0.000172  
                             &  38.26251        &   0.03407                 &  18.71$\pm$0.08   & 123.13$\pm$0.48  \\ 	
             2015 FK$_{37}$  & 189.38    & 65.90      & 0.973867 & 0.009071   & 156.048232  & 0.025643    & 171.942377       & 0.021302  
                             & 318.31574        &   0.10158                 &  35.82$\pm$17.18  &   5.66$\pm$2.84  \\ 	
             2016 FL$_{59}$  & 1237.17   & 124630.00  & 0.983370 & 1.684900   &  6.920160   & 0.002326    & 275.051954       & 2.225600  
                             &   1.48279        &  23.42600                 &     21$\pm$550770 &   2406$\pm$8061163 \\
             2017 CW$_{32}$  & 185.64    &  0.25      & 0.984069 & 0.000021   & 152.435852  & 0.000054    & 332.607656       & 0.000052  
                             & 180.77294        &   0.00015                 &  2.958$\pm$0.006  & 366.27$\pm$0.50  \\ 	
          \hline
        \end{tabular}
        \label{wetnos}
      \end{table*}
%
%

  \section{Including extreme Centaurs and comets}
\begin{figure}
        \centering
         \includegraphics[width=\linewidth]{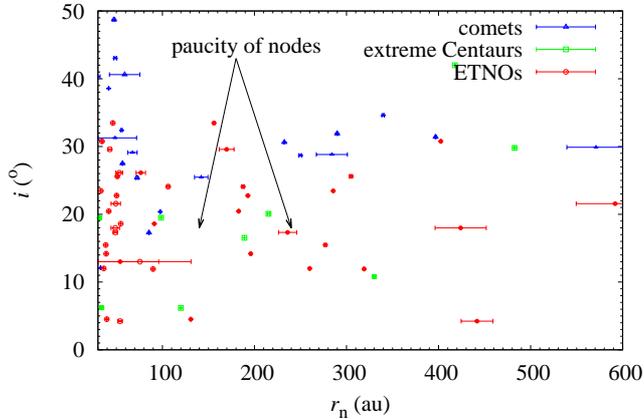}
         \caption{Orbital inclination as a function of the nodal distances for ETNOs (red circles, empty is $r_{-}$), extreme Centaurs 
                  (green squares, empty is $r_{-}$) and comets (blue triangles, empty is $r_{-}$). Nodal distances and error bars have been 
                  computed using heliocentric elements and uncertainties from Table~\ref{etnos}. 
                 }
         \label{inodesALL}
      \end{figure}

     As of 2017 June 5, the number of extreme Centaurs stands at 24 although the orbits of three of them are rather poor (see 
     Table~\ref{wetnos}); the number of comets with $a>150$~au is 389. If we plot a figure analogue to Fig.~\ref{inodes} but including data 
     for large semimajor axis Centaurs and comets with $a>150$~au as well, we obtain Fig.~\ref{inodesALL} that shows that the extreme 
     Centaurs follow the trend identified in the previous section. However, the distribution of the comets appears to be somewhat different 
     as their overall inclinations are higher than those of ETNOs and extreme Centaurs; most comets have parameters outside the displayed 
     range. This could be signalling a rather different dynamical origin for these comets when compared to ETNOs and extreme Centaurs. In 
     Fig.~\ref{inodesALL}, we have used heliocentric elements to compute the nodal distances applying Equation~(\ref{nodeseq}) as described 
     in the previous section, but we have removed all the objects with relative errors in the value of the semimajor axis greater than 50 
     per cent (i.e. 2003~SS$_{422}$, 2014~FE$_{72}$, 2006~UL$_{321}$, 2012~KA$_{51}$ and 2016 FL$_{59}$ have not been included). The 
     descending nodes of prograde extreme Centaurs tend to be systematically located near perihelion as expected of objects with argument of 
     perihelion preferentially close to 180\degr (see Table~\ref{wetnos}); this is exactly opposite to the trend observed for the ETNOs. 
     Many extreme Centaurs follow orbits nearly perpendicular to the ecliptic (see Table~\ref{wetnos}, this is why they do not appear in 
     Fig.~\ref{inodesALL}), which is one of the evolutionary tracks predicted by numerical simulations when a massive perturber is present 
     (de la Fuente Marcos et al. 2015, 2016; Batygin \& Brown 2016b; Brown \& Batygin 2016). Some extreme Centaurs may be former ETNOs. 
     However, the orbital diffussion scenario recently backed by Bannister et al. (2017) may not be able to reproduce this evolutionary 
     track.
%
%
%
%

  \section{Statistical significance}
     In the previous two sections, we have pointed out an interesting trend that is followed by both ETNOs and extreme Centaurs, that nodal 
     distances in the range (200, 250)~au seem to be avoided and the few present have associated values of the inclination within a rather 
     narrow range. This trend means that the distribution of the nodal distances of the ETNOs (and the extreme Centaurs) could be bimodal. 
     This tentative conclusion is based on data for 41 objects (20 ETNOs and 21 extreme Centaurs) as those with the poorest orbital 
     solutions have been excluded from the analysis. This proposed correlation is unlikely to be the result of observational bias as the 
     objects have been discovered by independent surveys, particularly the extreme Centaurs. If, as in the case of Jupiter, a 3:2 mean 
     motion resonance (now, with an unseen perturber) is being avoided by the nodes of the orbiting minor bodies, the value of the semimajor 
     axis of the putative planet should be close to 300~au or nearly 360~au if the resonance to be avoided is 2:1 instead of 3:2. If this 
     interpretation is correct, nodal distances in the range 200--300~au must be scarce for a combination of the three eccentric 
     populations.
%
%
      \begin{figure}
        \centering
         \includegraphics[width=\linewidth]{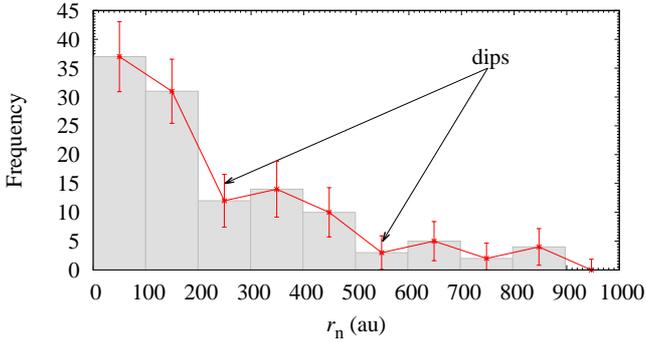}
         \caption{Similar to Fig.~\ref{jupc}, bottom panel, but using data from Tables~\ref{etnos} and \ref{wetnos}, and comets with 
                  $a>150$~au (IQR=244.5~au, $n$=118). The arrows point out the bins where the dips are located (see the text for details).
                 }
         \label{EW}
      \end{figure}
%
%

     Fig.~\ref{EW} is equivalent to Fig.~\ref{jupc}, bottom panel, but using data for ETNOs, extreme Centaurs and comets with $a>150$~au; 
     for a given object, only the node located farthest from the Sun has been counted. Although based on small-number statistics, the 
     difference between the bin centred at 150~au and the one at 250~au is over 4$\sigma$; in addition, the excess of the bin centred at 
     350~au with respect to the one at 550~au is about 3.7$\sigma$ ---in both cases using the $\sigma$-value at the dip. In principle, this 
     can be seen as an additional argument in support of our interpretation, if the dips are real. The second dip could be linked to a 
     second perturber or perhaps both dips signal the nodes of a single perturber.
      
  \section{Discussion}
     Although the features pointed out in the previous section (and the associated bimodality in nodal distances) seem to be real, it can 
     be disputed whether any dips present in Fig.~\ref{EW} are nearly entirely due to lack of completeness of the sample or mainly due to 
     observational biases. We believe that observational biases may not play a significant role in this case in light of the analysis 
     performed on the nodal distance versus inclination correlation, but more data are needed to find out a definite answer to that 
     question. 

     As for the presence of a planet at 300--400~au, the scenario explored by de Le{\'o}n, de la Fuente Marcos, \& de la Fuente Marcos 
     (2017) also places that range in semimajor axis within the region of interest where a perturber may have been able to disrupt the pair 
     of ETNOs (474640) 2004~VN$_{112}$--2013~RF$_{98}$, but only if its mass is in the range 10--20~$M_{\oplus}$. A closer perturber is also 
     advocated by Holman \& Payne (2016). Kenyon \& Bromley (2015, 2016) have shown that super-Earths may form at 125--750~au from the Sun.
  
  \section{Conclusions}
     In this Letter, we have documented a previously unnoticed correlation between the nodal distances and the inclinations of the ETNOs. 
     Although the size of the sample is small (22 ETNOs), the trend is also observed in other, perhaps related populations. The use of a 
     dynamical analogy with Jupiter leads us to perform a tentative interpretation of our findings. Our conclusions are as follows: 
     \begin{enumerate}[(i)]
        \item The distribution of the nodal distances of observed comets with $a<1\,000$~au is bimodal. This is a confirmation of results 
              obtained by Rickman et al. (2001). 
        \item We found strong evidence for a possible bimodal distribution of the nodal distances of the ETNOs in the form of a correlation 
              between their nodal distances and inclinations. 
        \item If the bimodal distribution in nodal distance observed in the case of Jupiter and those comets with $a<1\,000$~au is used
              as a dynamical analogue for the ETNOs, one trans-Plutonian planet with semimajor axis in the range 300--400~au may exist. 
              This result is consistent with the available data on both large semimajor axis Centaurs and comets. 
     \end{enumerate}

  \section*{Acknowledgements}
     We thank the anonymous referee for constructive, helpful, detailed and timely reports, S.~J. Aarseth, J. de Le\'on, J.-M. Petit, M.~T. 
     Bannister, D.~P. Whitmire, G. Carraro, E. Costa, D. Fabrycky, A.~V. Tutukov, S. Mashchenko, S. Deen and J. Higley for comments on ETNOs 
     and trans-Plutonian planets, and A.~I. G\'omez de Castro, I. Lizasoain and L. Hern\'andez Y\'a\~nez of the Universidad Complutense de 
     Madrid (UCM) for providing access to computing facilities. This work was partially supported by the Spanish `Ministerio de 
     Econom\'{\i}a y Competitividad' (MINECO) under grant ESP2014-54243-R. Part of the calculations and the data analysis were completed on 
     the EOLO cluster of the UCM, and we thank S. Cano Als\'ua for his help during this stage. EOLO, the HPC of Climate Change of the 
     International Campus of Excellence of Moncloa, is funded by the MECD and MICINN. This is a contribution to the CEI Moncloa. In 
     preparation of this Letter, we made use of the NASA Astrophysics Data System, the ASTRO-PH e-print server, and the MPC data server.

  \bsp
  \label{lastpage}
\end{document}